\documentclass[default,sn-mathphys-num,iicol]{sn-jnl}

\usepackage{tabularx}
\usepackage{graphicx}%
\usepackage{multirow}%
\usepackage{amsmath,amssymb,amsfonts}%
\usepackage{amsthm}%
\usepackage{mathrsfs}%
\usepackage[title]{appendix}%
\usepackage{xcolor}%
\usepackage{textcomp}%
\usepackage{manyfoot}%
\usepackage{booktabs}%
\usepackage{algorithm}%
\usepackage{algorithmicx}%
\usepackage{algpseudocode}%
\usepackage{listings}%
\usepackage{comment}
\usepackage{subfig}

\theoremstyle{thmstyleone}%
\theoremstyle{thmstyletwo}%

\theoremstyle{thmstylethree}%

\newcommand{\Th}{$^{232}$Th}
\newcommand{\U}{$^{238}$U}

\newcommand{\Tl}{$^{208}$Tl}
\newcommand{\A}{$\alpha$}
\newcommand{\B}{$\beta$}
\newcommand{\G}{$\gamma$}
\newcommand{\BG}{$\beta/\gamma$}

\newcommand{\Am}{$^{241}$Am}
\newcommand{\K}{$^{40}$K}
\newcommand{\gagg}{Gd$_3$Al$_2$Ga$_3$O$_{12}$}

\newcommand{\Eg}{E$_{\gamma}$}
\newcommand{\dt}{$\Delta t$}

\begin{document}

\title[GAGG]{Characterization of a GAGG detector for neutron measurements in underground laboratories}


\author[1,3]{\fnm{L.} \sur{Ascenzo}}\email{lorenzo.ascenzo@student.univaq.it}

\author[2,3]{\fnm{G.} \sur{Benato}}\email{giovanni.benato@gssi.it}

\author[2,3]{\fnm{Y.} \sur{Chu}}\email{yingjie.chu@gssi.it}

\author[3]{\fnm{G.} \sur{Di Carlo}}\email{giuseppe.dicarlo@lngs.infn.it}

\author[4,5]{\fnm{A.} \sur{Molinario}}\email{andrea.molinario@inaf.it}

\author[4,5]{\fnm{S.} \sur{Vernetto}}\email{vernetto@to.infn.it}

\affil[1]{\orgname{Universit\`a degli Studi dell'Aquila}, \orgaddress{\street{Via Vetoio 42}, \city{L'Aquila}, \postcode{67100}, \country{Italy}}}

\affil[2]{\orgname{Gran Sasso Science Institute}, \orgaddress{\street{Viale F. Crispi 7}, \city{L'Aquila}, \postcode{67100}, \country{Italy}}}

\affil[3]{\orgname{INFN - Laboratori Nazionali del Gran Sasso}, \orgaddress{\street{Via G. Acitelli 22} \city{Assergi (AQ)}, \postcode{67100}, \country{Italy}}}

\affil[4]{\orgname{INAF - Osservatorio Astrofisico di Torino}, \orgaddress{\street{Via Osservatorio 20}, \city{Pino Torinese (TO)}, \postcode{10025}, \country{Italy}}}

\affil[5]{\orgname{INFN - Sezione di Torino}, \orgaddress{\street{Via P. Giuria 1}, \city{Torino}, \postcode{10125}, \country{Italy}}}


\abstract{
  In rare events experiments, such as those devoted to the direct search of dark matter,
  a precise knowledge of the environmental gamma and neutron backgrounds
  is crucial for reaching the design experiment sensitivity.
  The neutron component is often poorly known due to the lack of a scalable detector technology
  for the precise measurement of low-flux neutron spectra.
  \gagg\ (GAGG) is a newly developed, high-density scintillating crystal with a high gadolinium content,
  which could allow  to exploit the high $(n,\gamma)$ cross section of $^{155}$Gd and $^{157}$Gd
  for neutron measurements in underground environments.
  GAGG crystals feature a high scintillation light yield, good timing performance,
  and the capability of particle identification via pulse-shape discrimination.
  In a low-background environment, the distinctive signature produced by neutron capture on gadolinium,
  namely a \BG\ cascade releasing up to 9\,MeV of total energy,
  and the efficient particle identification provided by GAGG could yield a background-free neutron capture signal.
  In this work, we present the characterization of a first GAGG detector prototype
  in terms of particle discrimination performance, intrinsic radioactive contamination, and neutron response.
}




\maketitle

\section{Introduction}\label{sec:intro}

In the field of rare events searches, a precise knowledge of the environmental backgrounds
is fundamental for the design of appropriate shieldings for the experiments.
Among the possible background sources, neutrons are often poorly known due to the lack of an
affordable detector technology for the measurement of low-flux neutron spectra from
thermal energies to 10\,MeV.

The last decade has witnessed the development of several garnet scintillation crystals
containing gadolinium, which has the two isotopes  -- $^{155}$Gd and $^{157}$Gd -- with the highest neutron capture cross section.
These crystals are now available with dimensions suitable for the use as particle detectors.
GAGG is of particular interest for its high scintillation
light yield, good timing performance, high density ($6.6$\,g/cm$^3$)
and capability of particle identification via pulse-shape discrimination~\cite{kamada,Kobayashi:2012az,iwanowska,tamagawa}.
In a low cosmic-ray environment, e.g. an underground laboratory, the distinctive signature produced by
neutron capture on gadolinium, namely a \G-ray cascade releasing $\sim$9\,MeV of total energy,
and the efficient particle identification provided by GAGG would yield a background-free
neutron signal.

Thanks to its 50.9\% gadolinium mass fraction, the application of GAGG as a
neutron detector is being considered by the scientific community~\cite{taggart,tyagi,fedorov2020,fedorov2024,wang2020}.
However, all tests performed so far have involved small-size GAGG crystals,
with a volume of up to a few cm$^3$, hence the considered neutron signature consisted of
X-rays and low-energy \G-rays from Gd de-excitation.
Exploiting the recent availability of crystals of $\sim$100\,cm$^3$ volume,
and the most up-to-date developments on Monte Carlo simulations of gamma ray cascades produced
by neutron capture reactions~\cite{Litaize:2015rco,AlmazanMolina:2019aoc,nudex},
we aim at demonstrating the possibility of detecting directly in the GAGG crystal
the high-energy \G-rays following the neutron capture on $^{155}$Gd and $^{157}$Gd.
In this perspective, the neutron signature is a \B/\G\ event in the range between 2.6\,MeV,
corresponding to the endpoint of natural \G\ radioactivity, and $\sim$9\,MeV,
corresponding to the endpoint of the \G\ cascades from the Gd isotopes.
When operated in Bonner spheres, such a detector would allow to perform neutron spectroscopy in low-background environments,
and could represent a valid alternative to the well-established $^3$He proportional counters,
whose price and availability on the market have fluctuated significantly over the last decades.

In this article, we present the characterization of a first detector prototype,
consisting of a $\sim$100\,cm$^3$ GAGG crystal, read-out with a photomultiplier (PMT),
operated underground in the Hall A of the Gran Sasso National Laboratory (LNGS) of INFN, Italy,
between January and October, 2024.
In Sec.~\ref{sec:det} we describe the experimental setup and the measurements performed.
In Sec.~\ref{sec:data} we detail the data processing and detector performance.
In Sec.~\ref{sec:bkg} and~\ref{sec:neutrons} we report on the background characterization
and on the source and environmental neutron measurements.
Finally, in Sec.~\ref{sec:conclusion} we provide a plan for the future developments of the detector.

\section{Detector and data taking}\label{sec:det}

\begin{figure}[htbp]
  \centering
  \includegraphics[height=0.227\textheight]{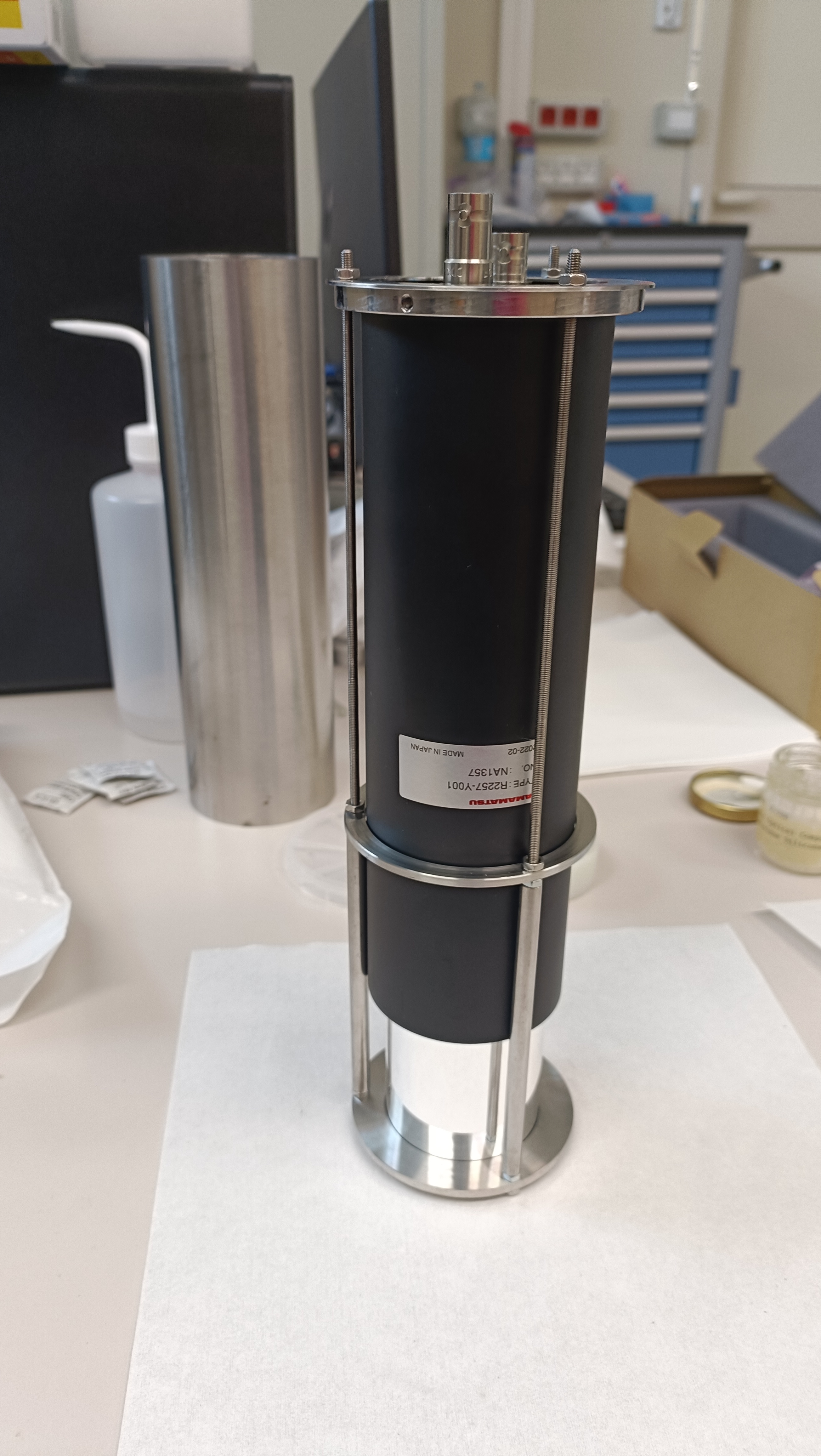}\quad
  \includegraphics[height=0.227\textheight]{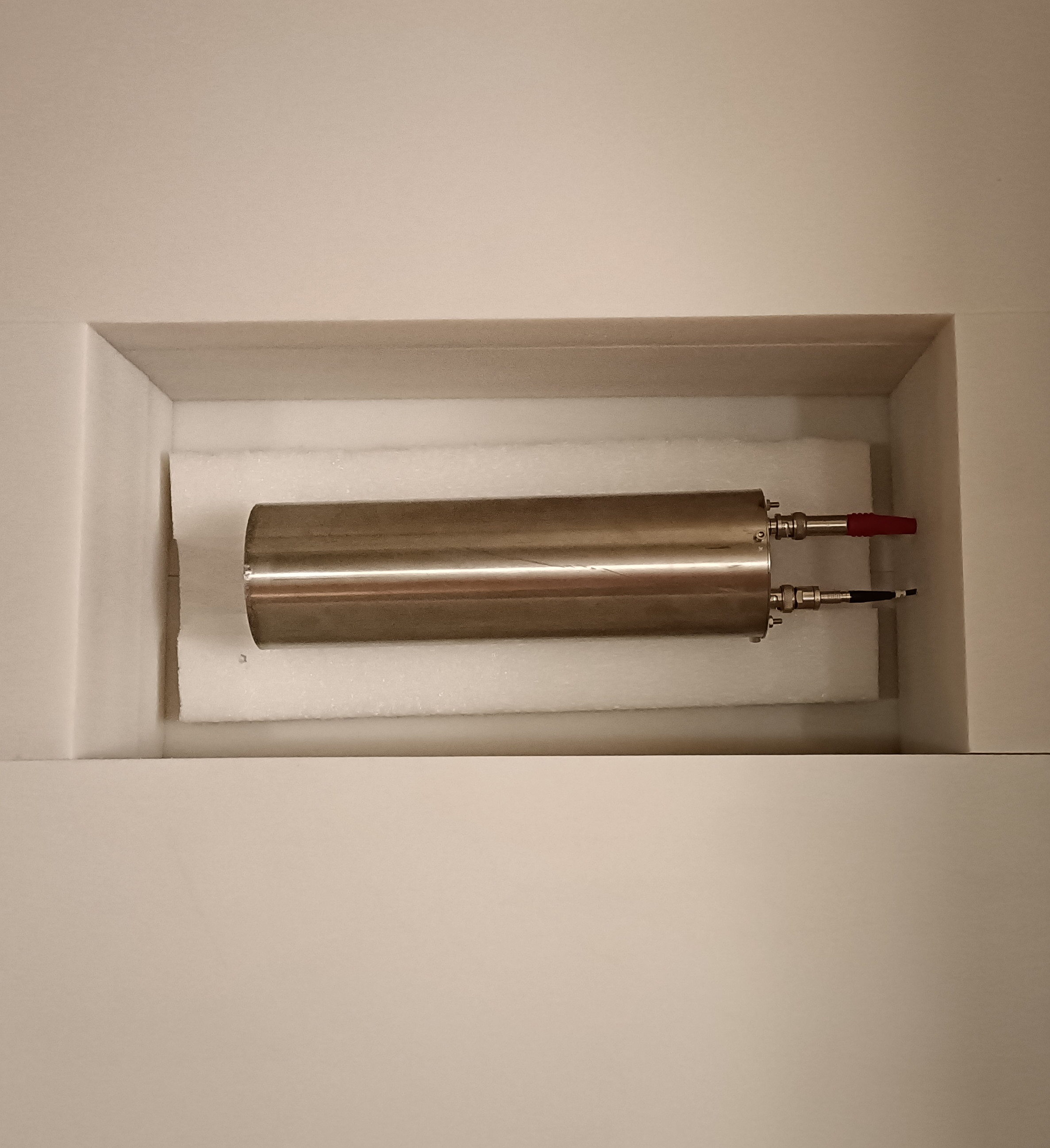}
  \caption{Left: the detector during assembly, with the PMT on the top and the GAGG crystal
    wrapped in reflector foil in the bottom.
    Right: the detector enclosed in its stainless steel case and installed in the borated PE shielding.}\label{fig:detector}
\end{figure}

The detector prototype is composed of a cylindrical GAGG:Ce crystal with 5\,cm diameter and 5\,cm height,
corresponding to a mass of 0.65\,kg, coupled to a Hamamatsu R2257 PMT and mounted in a custom-made stainless steel case (Fig.~\ref{fig:detector}).
To maximize the light collection, the crystal is wrapped in Vikuiti ESR reflector foil.
The PMT is supplied with $-1.4$\,kV using a CAEN V6533N high-voltage module,
and 2\,\textmu s long signal waveforms are read-out at 250\,MHz using a 14-bit CAEN V1725 digitizer.
The GAGG scintillation signal features a fast decay constant of $\sim$90\,ns,
and a slow one of $\sim$400\,ns, with the ratio of the fast and slow component
depending on the interacting particle (see later, Fig.~\ref{fig:ap}).

For the purpose of measuring \G\ events above 2.6\,MeV,
it is crucial to operate the detector underground to bypass the otherwise overwhelming background from cosmic muons,
and to discriminate them from \A\ events which could represent a background in our region of interest.
Our choice of sampling time and window length
is thus motivated by the need to distinguish \A\ from \B/\G\ particles via pulse-shape discrimination.

We performed several measurements aimed at a full characterization of the detector response and background,
and at a first evaluation of its performance as a neutron detector.
We started with \G\ calibrations using \Th, \U\ and $^{241}$Am sources,
then placed the detector in a 20\,cm thick polyethylene (PE) shielding with 5\% boron loading,
depicted in Fig.~\ref{fig:detector},
and acquired background data for $\sim$2.5 months.
We then performed two calibration runs with an AmBe neutron source inside the shielding:
in the first run we placed a 5\,cm thick PE moderator between the source and the detector;
in a second run we removed the PE moderator, and placed a 5\,cm thick Cu layer
between the source and the detector to shield the 4.4\,MeV \G-ray produced by the AmBe source itself.
In addition, we acquired data for one month with the completely unshielded detector
to measure the thermal neutron flux in Hall A of LNGS.
Finally, we acquired again backgrund data for $\sim$1 month using a 400\,ns acquisition window
for a more precise evaluation of the \Th\ contamination in the crystal,
as detailed in Sec.~\ref{sec:bkg}.
A summary of the performed measurements is given in Tab.~\ref{tab:data}.

\begin{table}[htbp]
  \centering
  \caption{List of measurements performed with the GAGG prototype detector.}\label{tab:data}
  \begin{tabular}{lc}
    \toprule
    Measurement description & Duration \\
    \midrule
    $^{232}$Th calibration            &  \\
    $^{238}$U calibration            &  \\
    $^{241}$Am calibration            &  \\
    Background (with 20\,cm borated PE) & 77.5\,days \\
    AmBe with 5\,cm PE               & 21.5\,hr \\
    AmBe with 5\,cm Cu \G\ shield    & 22.5\,hr \\
    Environmental (thermal) neutrons  & 31.6\,days \\
    Background with 400\,ns window & 27\,days\\
    \bottomrule
  \end{tabular}
\end{table}

\section{Data processing}\label{sec:data}

\begin{figure}[htbp]
  \centering
  \includegraphics[width=\columnwidth]{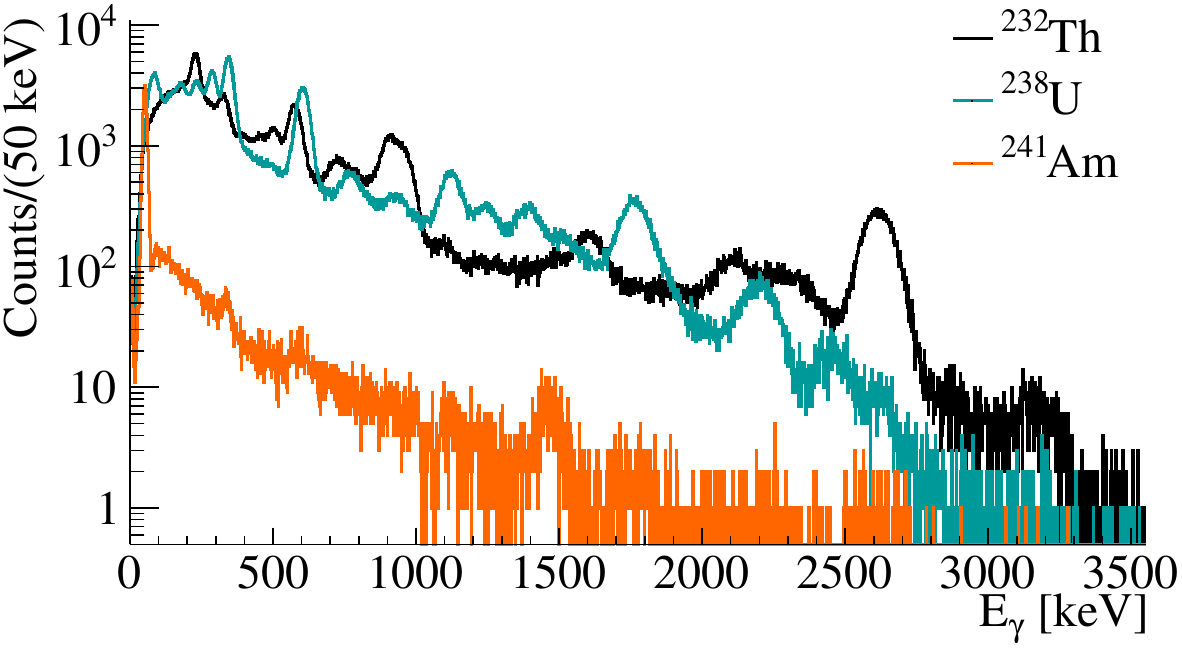}
  \caption{Calibration spectra recorded with \Th\ (black), \U\ (teal) and \Am\ (orange) sources
    placed in front of the detector. The only visible signature produced by \Am\
    is the 59.5\,keV peak reconstructed in the very left of the spectrum.
    The orange continuum is induced by the internal and environmental \BG\ background.}\label{fig:calibration}
\end{figure}

\begin{figure}[htbp]
  \centering
  \includegraphics[width=\columnwidth]{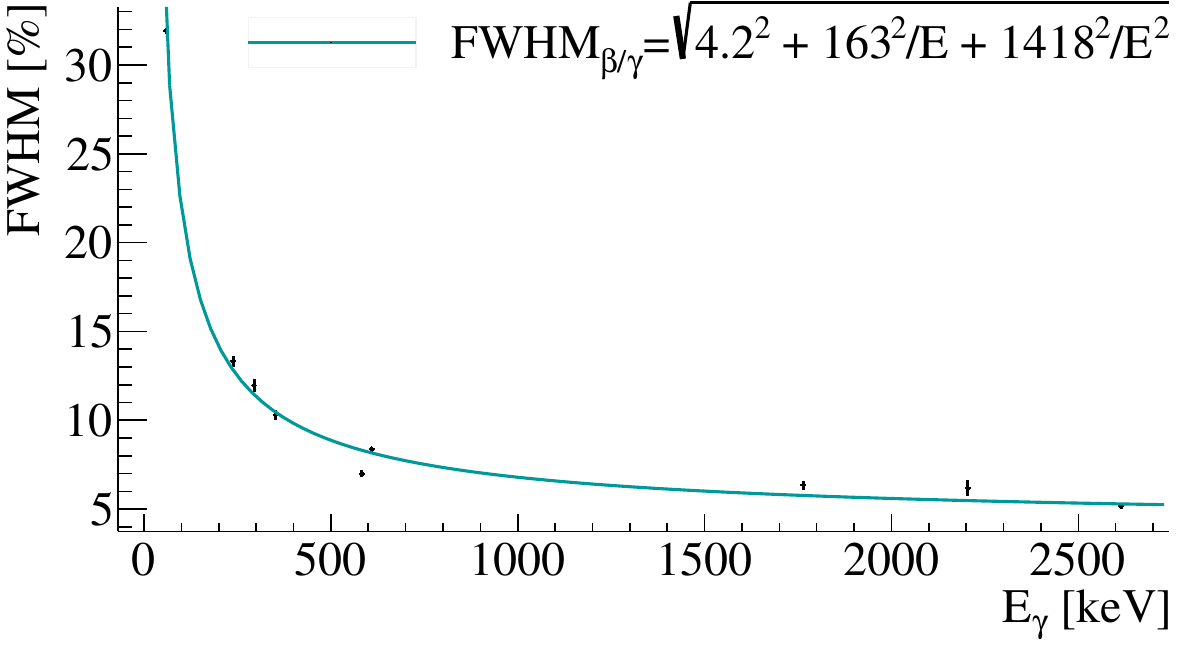}
  \caption{Resolution curve obtained from the most prominent peaks visible in the \Th, \U, and \Am\ spectra.}\label{fig:resolution}
\end{figure}

We process the data using the Octopus software~\cite{octopus},
originally developed for the reconstruction of data from bolometric detectors
and further expanded for analyzing the signals from the GAGG detector.
We evaluate the energy of each triggered event by integrating the waveform,
and reject events with a trigger shifted from its predefined $0.45$\,\textmu s position in the waveform
as well as events with an unstable baseline.
The combination of these two cuts remove $\sim1\%$ of the events.
We show in Fig.~\ref{fig:calibration} the calibration spectra collected with
\U, \Th\ and $^{241}$Am sources placed in front of the detector.
We fit the most prominent peaks in the calibration spectra, and extracted the resolution curve
reported in Fig.~\ref{fig:resolution}, indicating an asymptotic relative full width at half maximum (FWHM) of $\sim4.2\%$
in the region of interest for the Gd de-excitation \G\ rays, i.e. above 3\,MeV.

\begin{figure}[htbp]
  \centering
  \includegraphics[width=\columnwidth]{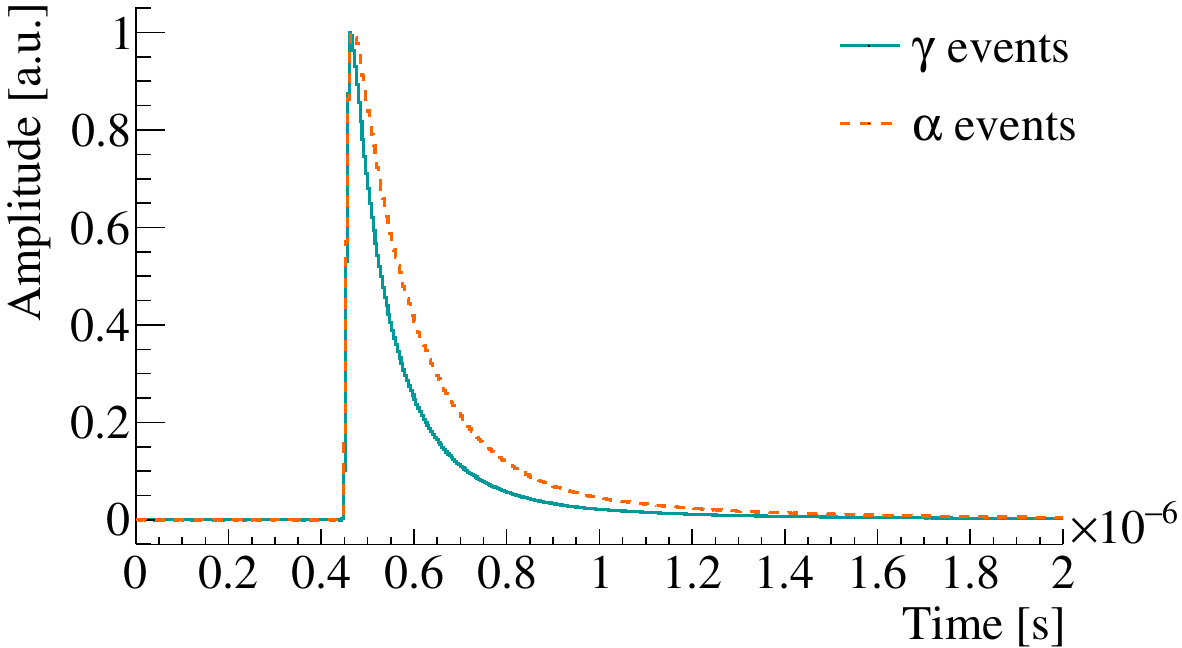}\\
  \includegraphics[width=\columnwidth]{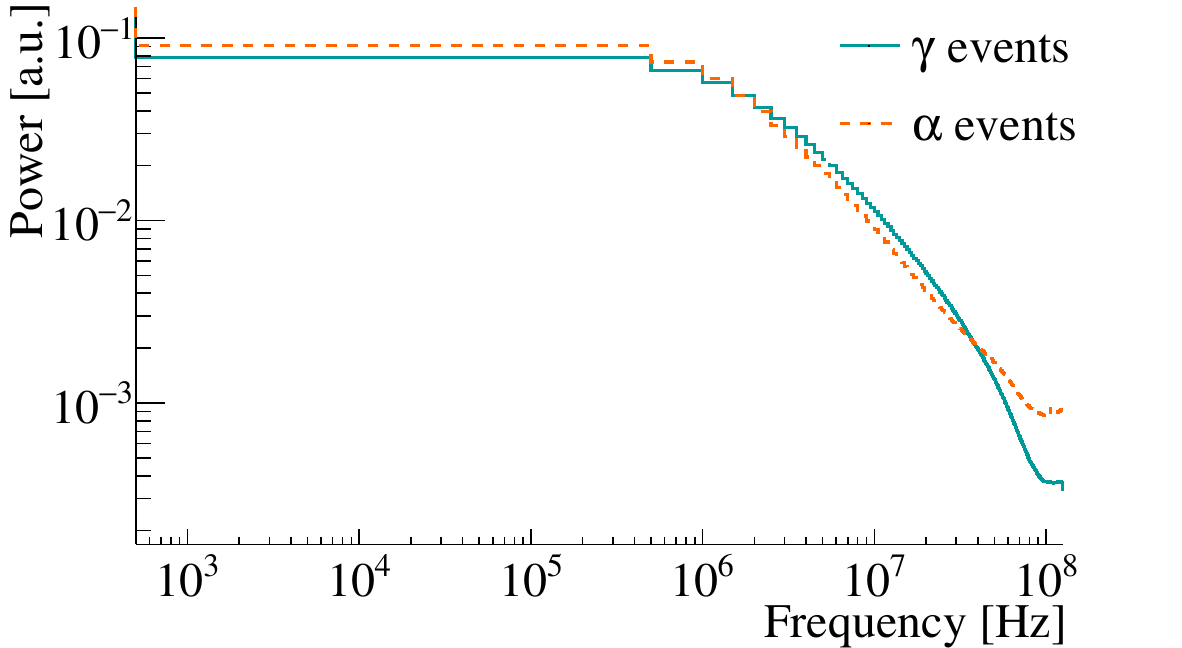}
  \caption{Normalized average pulse (top) and average power spectrum (bottom) for \BG\ and \A\ events.}\label{fig:ap}
\end{figure}

The discrimination between \BG\ and \A\ events is based on the different
pulse shape of the two event types, which is evident from Fig.~\ref{fig:ap} (top).
Rather than exploiting the commonly used mean-decay-time algorithm~\cite{Lee:2015iaa},
which is sensitive to the trigger position within the digitized window,
we opted to develop a pulse-shape discrimination (PSD) algorithm
using the events power spectrum, as shown in the bottom of Fig.~\ref{fig:ap}, thus completely bypassing any possible misalignment
of the events within the acquisition window.
Specifically, we compute the $\chi^2$ between the power spectrum of each single event $w_f$
and the average power spectrum $\bar{w}_f$ of \G\ events:
\begin{equation}\label{eq:chi2}
  \chi^2 = \sum_{f=0}^{f_{max}} \left( w_f - A_{\chi^2}\cdot\bar{w}_f \right)^2\quad,
\end{equation}
where $f$ indicates the frequency index of the discrete power spectrum,
and $A_{\chi^2}$ is a normalization factor against which $\chi^2$ is minimized.
Such a normalization factor will be proportional to the pulse integral $I$ (computed in the time domain)
for \BG\ events, and significantly deviate from it for \A\ events.

\begin{figure}[htbp]
  \centering
  \includegraphics[width=\columnwidth]{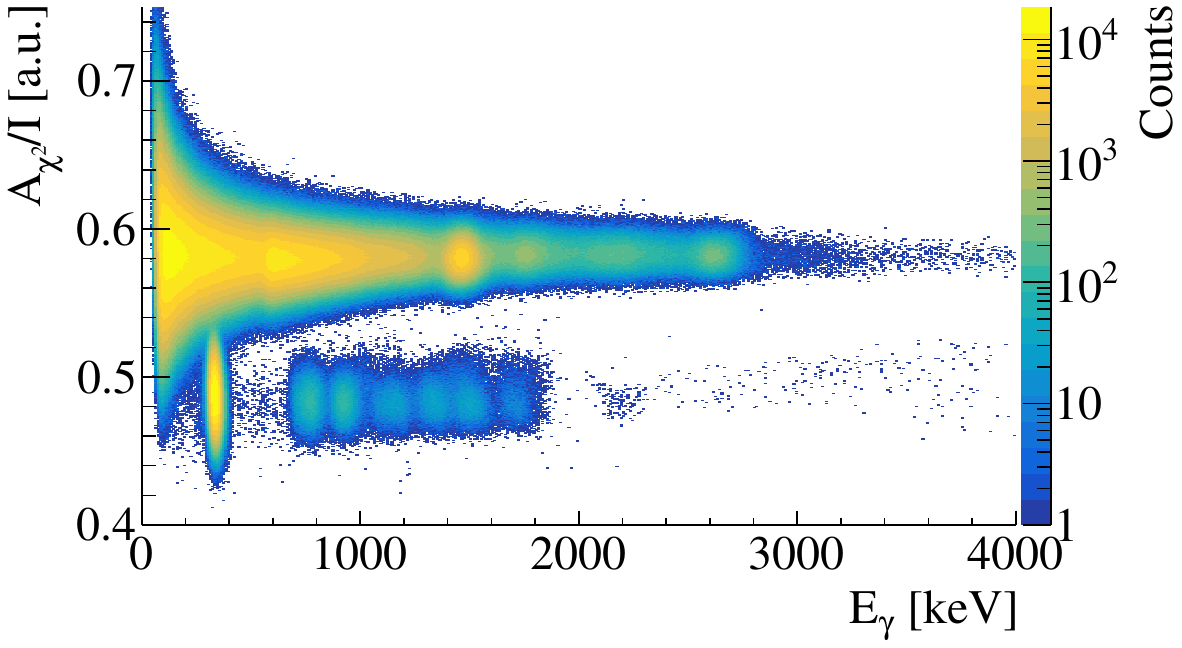}
  \caption{Distribution of the PSD variable $A_{\chi^2}/I$ as a function of energy for background data.
  The upper horizontal band corresponds to \BG\ events, the lower one to \A\ events.}\label{fig:psd}
\end{figure}

The distribution of the PSD variable $A_{\chi^2}/I$ as a function of energy is divided in two separate bands for the \BG\ and \A\ events, as shown in Fig.~\ref{fig:psd} for background data.
This figure also shows that the energy of \A\ events is quenched at the $\sim15\%$ level
with respect to \BG\ events, a value similar to what reported in literature~\cite{TAMAGAWA2015192}.

\section{Background characterization}\label{sec:bkg}

\begin{figure}[htbp]
  \centering
  \includegraphics[width=\columnwidth]{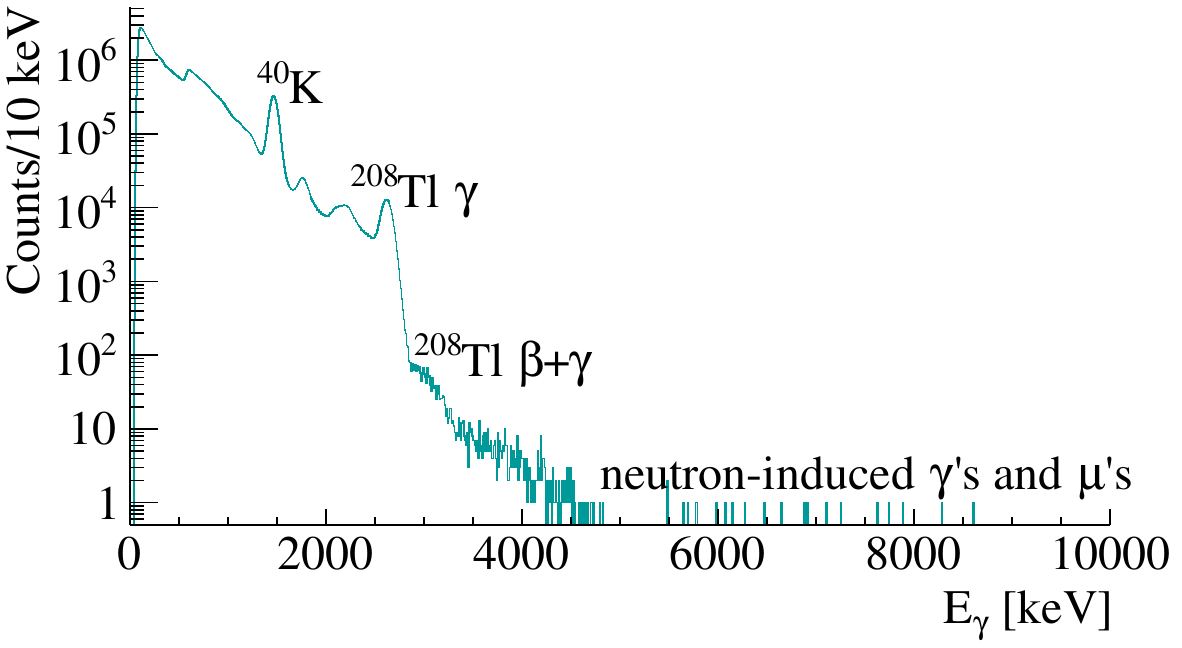}
  \caption{Background spectrum in the \BG\ band.
    The continuum between the $^{208}$Tl line and 5\,MeV can be ascribed to the pile-up of \B\ and \G\ decays from $^{208}$Tl.
    Events at energies above 5\,MeV are due to \G's emitted following neutron captures on Gd, and residual muons.}\label{fig:bkggamma}
\end{figure}

\begin{figure}[htbp]
  \centering
  \includegraphics[width=\columnwidth]{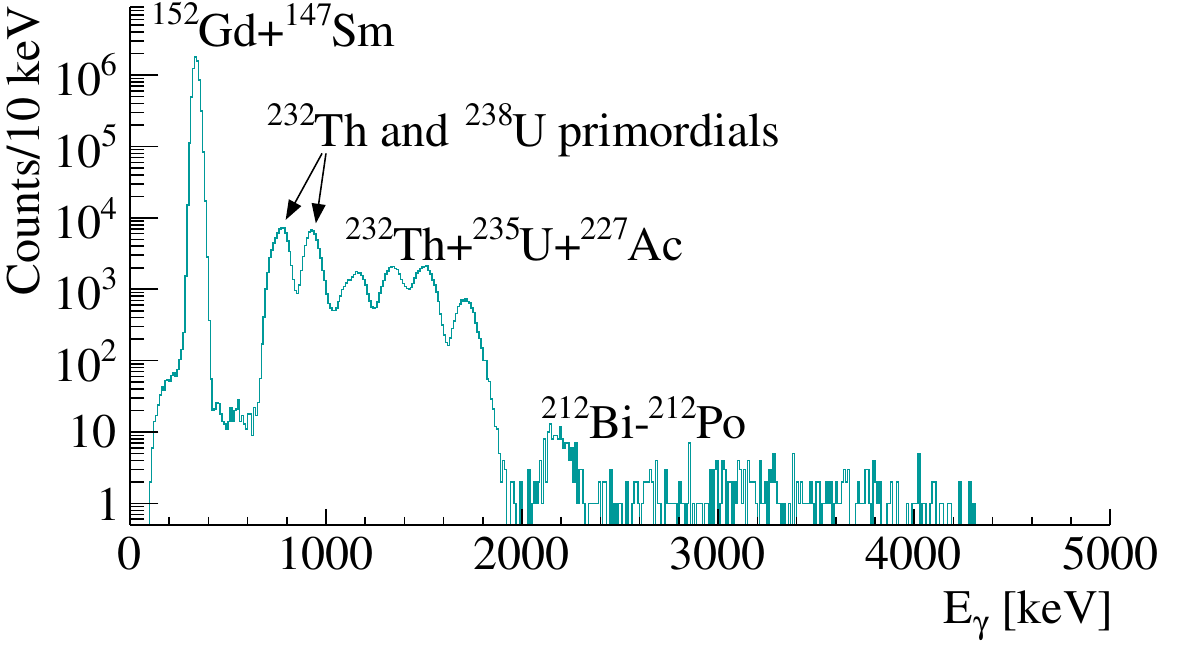}
  \caption{Background spectra in the \A\ band, here represented in the \G\ energy scale E$_{\gamma}$.
    The first peak is originated from the $^{152}$Gd and $^{147}$Sm \A\ decays.
    The six bumps in the central part comes from various \A\ decays from the $^{232}$Th, $^{238}$U, and $^{235}$U chains.
    The small peak at E$_{\gamma}$\,=\,2.2\,MeV and the sparse continuum up to 4.5\,MeV
    are due to the pile-up of $^{212}$Bi \BG\ and $^{212}$Po \A\ decays.}\label{fig:bkgalpha}
\end{figure}

We apply the PSD cut to the 77.5\,days of background data, obtaining the \BG\ and \A\ spectra
depicted in Figs.~\ref{fig:bkggamma} and~\ref{fig:bkgalpha}, respectively.
The \BG\ spectrum features two prominent \G\ lines, corresponding to the
1460\,keV peak from \K\ and to the 2615\,keV peak from \Tl.
The continuum between them is attributable to external \G's from the \Th\ and \U\ decay chains
undergoing Compton scattering in the crystal, and \B\ and \G's emitted and fully absorbed within the crystal itself.

A second, less intense continuum reaching 5\,MeV is visible:
we ascribe it to a \Th\ contamination in the crystal,
which eventually leads to the presence and decay of \Tl.
In fact, the signature from a \Tl\ decaying outside the crystal
would be a combination of mono-energetic \G\ lines at 2615\,keV and 583\,keV,
plus their sum and Compton continua.
On the other hand, the signature from a \Tl\ nucleus decaying inside the crystal
is a \B\ with energy up to 1.8\,MeV plus possibly the full absorption of the 583 and 2615\,keV \G.
The latter are emitted in coincidence in 85\% of the cases and have an overall probability of $\sim$50\%
to be fully contained in the GAGG crystal under study.
As a consequence, we expect a continuum reaching the $Q$-value of the \Tl\ decay, 5\,MeV.
To our knowledge, no other isotope -- either long-lived or belonging to a decay chain -- could induce such a \BG\ background.

Finally, 22 events with \Eg$>$5\,MeV are detected.
The expected number of events induced by muons crossing the crystal, which would cover a much wider energy range reaching $\sim70$\,MeV,
is $\lesssim$5.
Moreover, we expect environmental neutrons to be fully absorbed by the borated PE shield.
Hence, we estimate that $80\%$ of the events above 5\,MeV are induced by neutrons
produced e.g. by $(\alpha,n)$ reactions in the detector setup and subsequently captured by $^{155}$Gd or $^{157}$Gd.
Overall, the background in the $[5,10]$\,MeV region is ($0.28\pm0.06$) events/day.


The \A\ spectrum (Fig.~\ref{fig:bkgalpha}) features three main components.
A prominent peak is visible at roughly 240\,keV of the \G\ energy scale (\Eg), with a rate of 1\,event/s.
About half of these events can be attributed to $^{152}$Gd,
decaying \A\ with a $Q$-value of 2205\,keV. The other half are thought to be induced
by a crystal contamination of $^{147}$Sm, that decays \A\ with a $Q$-value of 2311\,keV.
$^{147}$Sm was indeed found to be a contaminant in two Gd samples measured via
Inductively Coupled Plasma Mass Spectrometry (ICP-MS)~\cite{nisi}.

A set of six bumps, some of which are clearly composed of overlapping components,
is present at \Eg$\in[700,1900]$\,keV.
After a careful analysis of the \A\ spectrum, which is detailed below,
we conclude that the leftmost two peaks are induced by the early parts of the \Th\ and \U\ decay chains,
and the other four correspond to various \A's emitted by the late part of the \Th\ chain,
the early part of $^{235}$U chain, and by a $^{227}$Ac contamination of the crystal.

A small peak is present at \Eg$\simeq$2200\,keV, followed by a sparse continuum reaching 4400\,keV.
This component is due to the pile-up of $^{212}$Bi and $^{212}$Po decay events from the \Th\ chain,
which are reconstructed within the same 2\,\textmu s acquisition windows
and are composed of an \A\ and a \BG\ event.
This interpretation is supported by the fact that the PSD distribution of these events (Fig.~\ref{fig:psd})
gets broader with energy and slowly drifts towards the \BG\ band.


The interpretation and energy calibration of the \A\ spectrum is particularly complicated
by the energy dependence of the \A\ quenching.
To identify the individual \A\ components, we search for time-delayed coincidences (TDC)
between subsequent \A\ events~\cite{Baccolo:2021odk,Azzolini:2021yft} belonging
to the $^{238}$U, $^{235}$U and $^{232}$Th decay chains, reported in Tab.~\ref{tab:dc}.
This analysis is sensitive to TDCs with half-life values ranging from the length of the signal window (2\,\textmu s)
to a few times the inverse of the event rate in the \A\ band, which is about 30\,s.
For much longer half-life values, the occurrence of random delayed coincidences would become dominant.
\begin{table}[htbp]
  \centering
  \caption{List of TDC belonging to the $^{238}$U, $^{235}$U and $^{232}$Th decay chains,
    with half-life values less than 5\,min.}\label{tab:dc}
  \begin{tabular}{cccc}
    \toprule
    Progenitor & Decay & Q-value & Half-life\\
               &       & [keV]   & [s] \\
    \midrule
    \multirow{3}{*}{$^{235}$U} & $^{223}$Ra$\rightarrow^{219}$Rn+\A & 5979 & \\
                               & $^{219}$Rn$\rightarrow^{215}$Po+\A & 6946 & 3.96\\
                               & $^{215}$Po$\rightarrow^{211}$Pb+\A & 7526 & 1.78$\cdot10^{-3}$\\
    \midrule
    \multirow{3}{*}{$^{232}$Th} & $^{224}$Ra$\rightarrow^{220}$Rn+\A & 5789 & \\
                                & $^{220}$Rn$\rightarrow^{216}$Po+\A & 6405 & 55.6 \\
                                & $^{216}$Po$\rightarrow^{212}$Pb+\A & 6906 & 0.1442 \\
    \midrule
    \multirow{2}{*}{$^{232}$Th} & $^{212}$Bi$\rightarrow^{212}$Po+\B & 2252 \\
                                & $^{212}$Po$\rightarrow^{208}$Pb+\A & 8954 & 2.99$\cdot10^{-7}$\\
    \midrule
    \multirow{2}{*}{$^{232}$Th} & $^{212}$Bi$\rightarrow^{208}$Tl+\A & 6207 \\
                                & $^{208}$Tl$\rightarrow^{208}$Pb+\B & 4999 & 183 \\
    \midrule
    \multirow{2}{*}{$^{238}$U} & $^{222}$Rn$\rightarrow^{218}$Po+\A & 5590 & \\
                               & $^{218}$Po$\rightarrow^{214}$Pb+\A & 6115 & 186 \\
    \midrule
    \multirow{2}{*}{$^{238}$U} & $^{214}$Bi$\rightarrow^{214}$Po+\B & 3269 & \\
                               & $^{214}$Po$\rightarrow^{210}$Pb+\A & 7834 & $1.64\cdot10^{-4}$ \\
    \bottomrule
  \end{tabular}
\end{table}


\begin{figure}[htbp]
\centering
\includegraphics[width=\columnwidth]{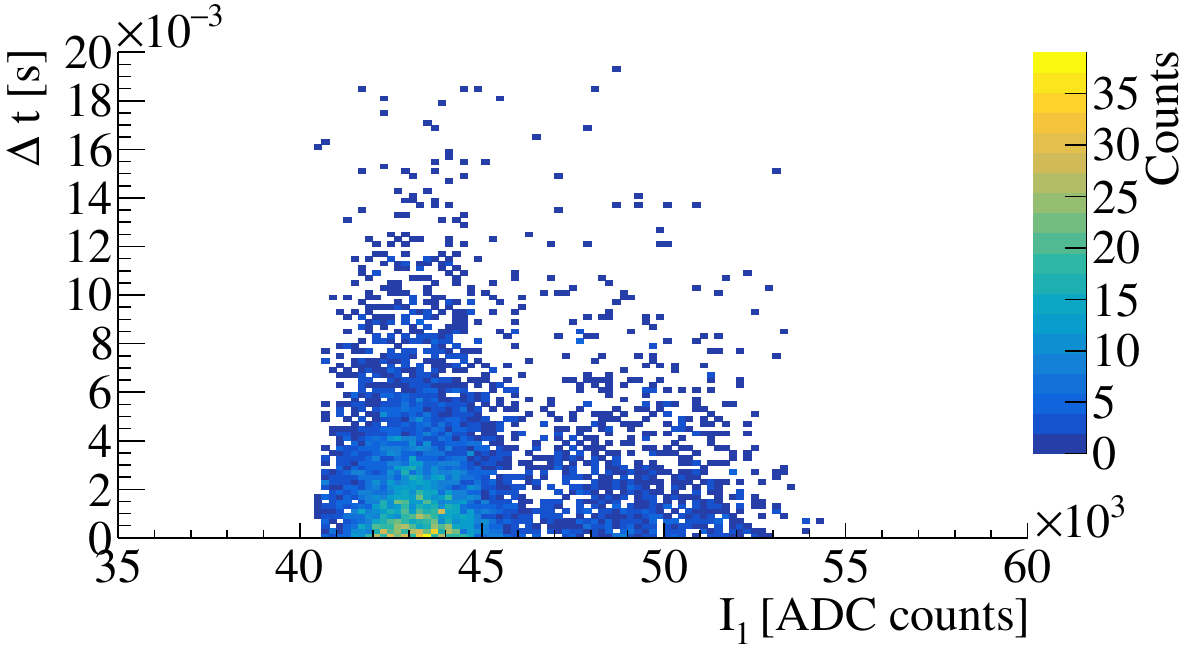}
\caption{Distribution of \dt\ between the $^{219}$Rn and $^{215}$Po \A\ decays vs. the pulse integral $I_1$ of $^{219}$Rn.
  A clear exponential decay in time is evident in $I_1\in[4,4.7]\cdot10^4$, indicating the presence of $^{219}$Rn.
  The other population at $I_1\in[4.7,5.3]\cdot10^4$ is due to $^{219}$Rn decays
  in which both \A\ and \G\ are emitted (11\% of the cases).}\label{fig:dtvsarea}
\end{figure}

\begin{figure}[htbp]
\centering
\includegraphics[width=\columnwidth]{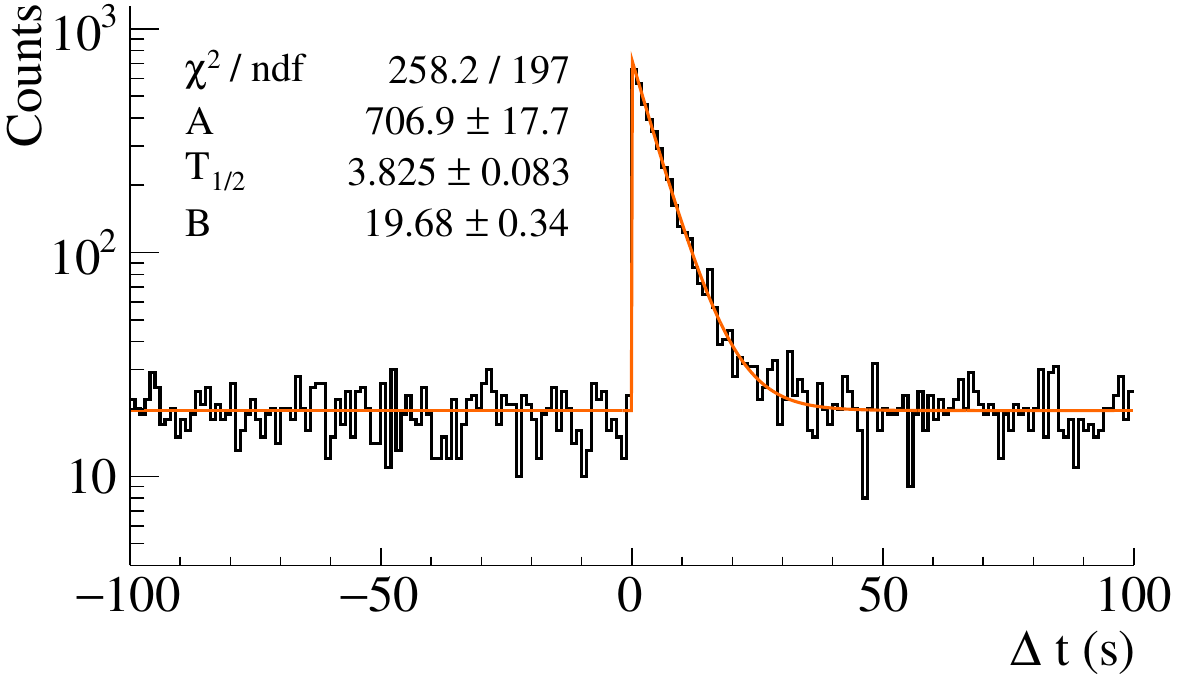}
\includegraphics[width=\columnwidth]{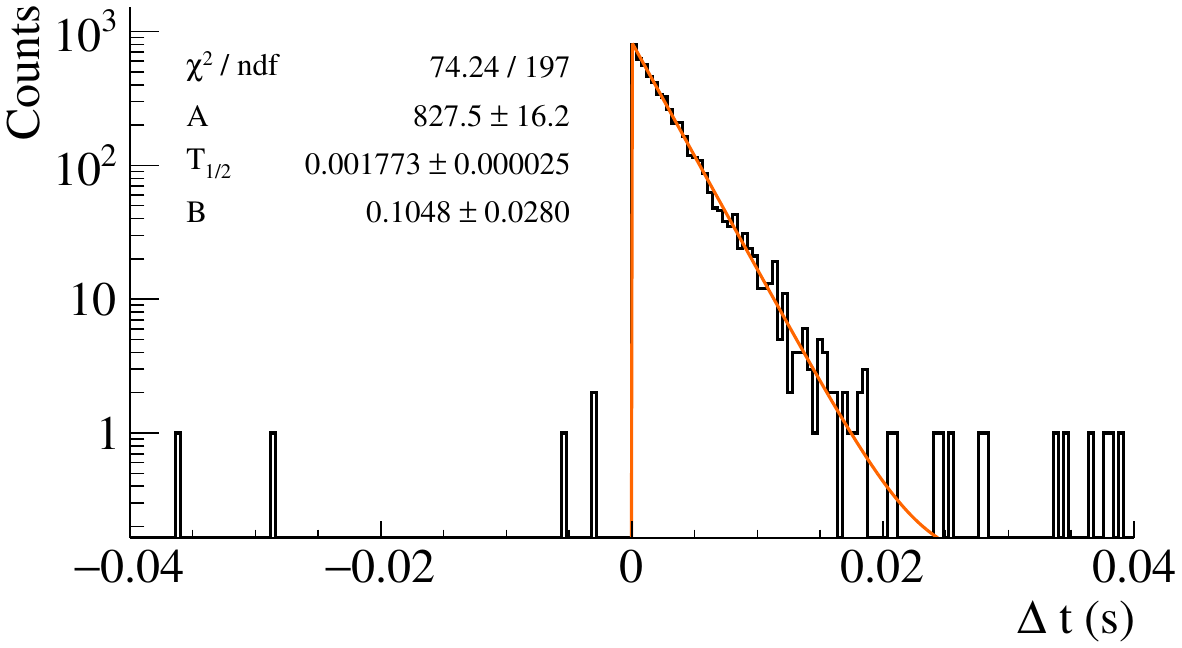}
\caption{Exponential fit to the \dt\ distribution for the $^{223}$Ra$\rightarrow^{219}$Rn (top)
  and $^{219}$Rn$\rightarrow^{215}$Po (bottom) \A\ decays in the $^{235}$U chain.
  The best-fit values are reported in the figures.}\label{fig:dt_example}
\end{figure}

For a first identification of the \A\ peaks, we search for the fastest \A-\A\ TDC,
corresponding to $^{219}$Rn$\rightarrow^{215}$Po$\rightarrow^{211}$Pb.
We compute the time delay \dt\ between subsequent \A\ events in a suitable range of pulse integrals $I$,
and plot it against the pulse integral of the first event in the pair, $I_1$.
If the $^{219}$Rn triplet is present, we expect a large number of events
in a well-defined range of pulse integral for \dt$<$10\,ms
(corresponding to about 5 times the half-life of $^{215}$Po).
The background from random delayed coincidences should be negligible in this time window.
Moreover, the $^{215}$Po would produce an exponentially decreasing distribution of \dt,
while random coincidences would only produce a flat distribution.
Figure~\ref{fig:dtvsarea} shows the presence of two populations centered around
$I_1=4.3\cdot10^4$ and $I_1=4.9\cdot10^4$, with an exponentially decaying \dt.
Both correspond to the $^{219}$Rn$\rightarrow^{215}$Po TDC,
with the first population composed of pure \A\ events,
and the second of \A\ events accompanied by a low-energy \G\ de-excitation
which significantly shifts $I_1$ to larger values thanks to the higher light-yield of \BG\ events.
We then reconstruct the \dt\ distribution between \A\ events
including events both backward and forward in time to have an accurate estimation of the background level.
We finally fit the resulting distribution with an exponential plus a flat background $B$:
$$
f(\Delta t)=A \exp\left(-\frac{\ln{2}\cdot\Delta t}{T_{1/2}}\right)+B\quad,
$$
where $T_{1/2}$ is the half-life of the second isotope of the doublet.
The same procedure is repeated to look for the other \A\ decay of the triplet.
The fits to the time delay distributions for $^{223}$Ra$\rightarrow^{219}$Rn
and $^{219}$Rn$\rightarrow^{215}$Po \A\ decays are shown in Fig.~\ref{fig:dt_example}.

We apply this method also to the second \A\ triplet, from the $^{232}$Th chain.
The measured half-life values of the decays are consistent with the expected ones for both \A\ triplets,
as reported in Tab.~\ref{tab:halflives}.
This is a robust confirmation of the correct identification of the \A\ decays.

We also search for the \A-\A\ TDC from $^{222}$Rn$\rightarrow^{218}$Po$\rightarrow^{214}$Pb
and the \B-\A\ TDC from $^{214}$Bi$\rightarrow^{214}$Po$\rightarrow^{210}$Pb,
both belonging to the $^{238}$U chain, finding no candidate event.
This indicates that the lower part of the $^{238}$U chain is not present in the GAGG crystal under study.
Similarly, we search for the $^{212}$Bi$\rightarrow^{208}$Tl$\rightarrow^{208}$Pb doublet from the \Th\ chain,
but find no excess of signal events over a very high background
induced by the \B\ candidate events.

\begin{table}
  \centering
  \caption{Half-life values reconstructed from the identified TDCs,
    compared with the expected values obtained from literature~\cite{livechart}.}\label{tab:halflives}
  \begin{tabular}{ccc}
    \toprule
    Isotope & Measured $T_{1/2}$ & Expected $T_{1/2}$ \\
    \midrule
    $^{219}$Rn & $3.83\pm0.08$\,s  & 3.96\,s \\
    $^{215}$Po & $1.77\pm0.03$\,ms & 1.78\,ms \\
    $^{220}$Rn & $55\pm1$\,s       & 56\,s \\
    $^{216}$Po & $141\pm2$\,ms     & 144\,ms \\
    \bottomrule
  \end{tabular}
\end{table}

\begin{figure}[htbp]
  \centering
  \includegraphics[width=\columnwidth]{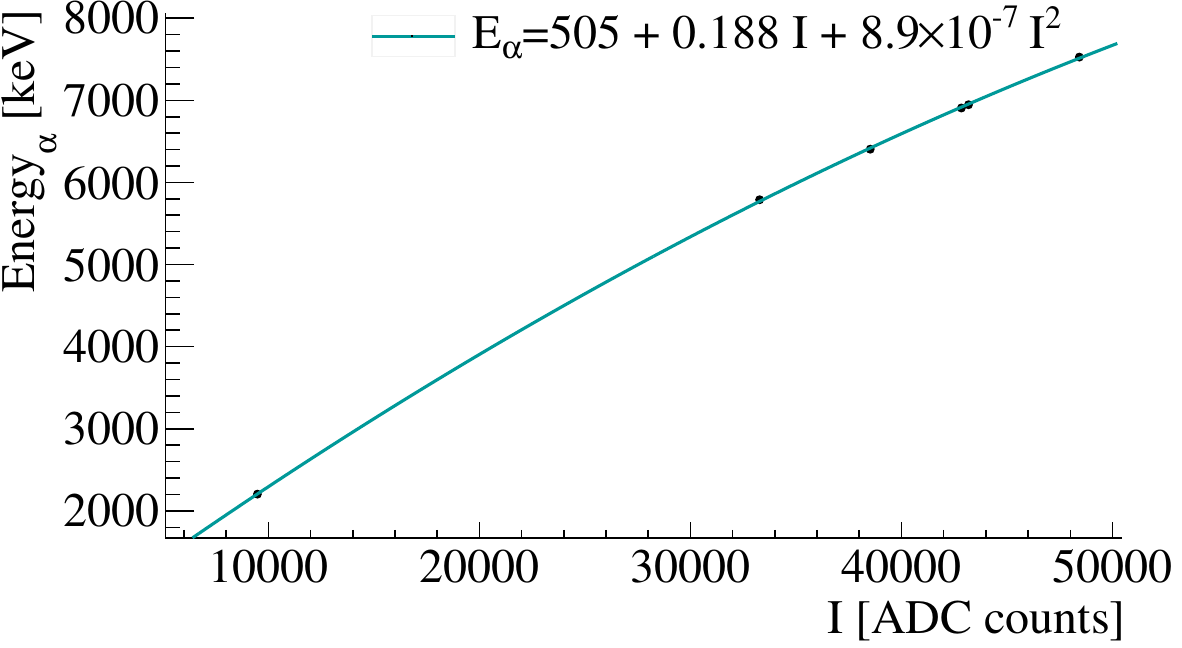}
  \caption{Polynomial fit to the \A\ energy, E$_{\alpha}$, vs. pulse integral $I$ of the \A\ peaks
    identified with the delayed coincidence study plus that from $^{152}$Gd.}\label{fig:alpha_calib}
\end{figure}

\begin{figure}[htbp]
  \centering
  \includegraphics[width=\columnwidth]{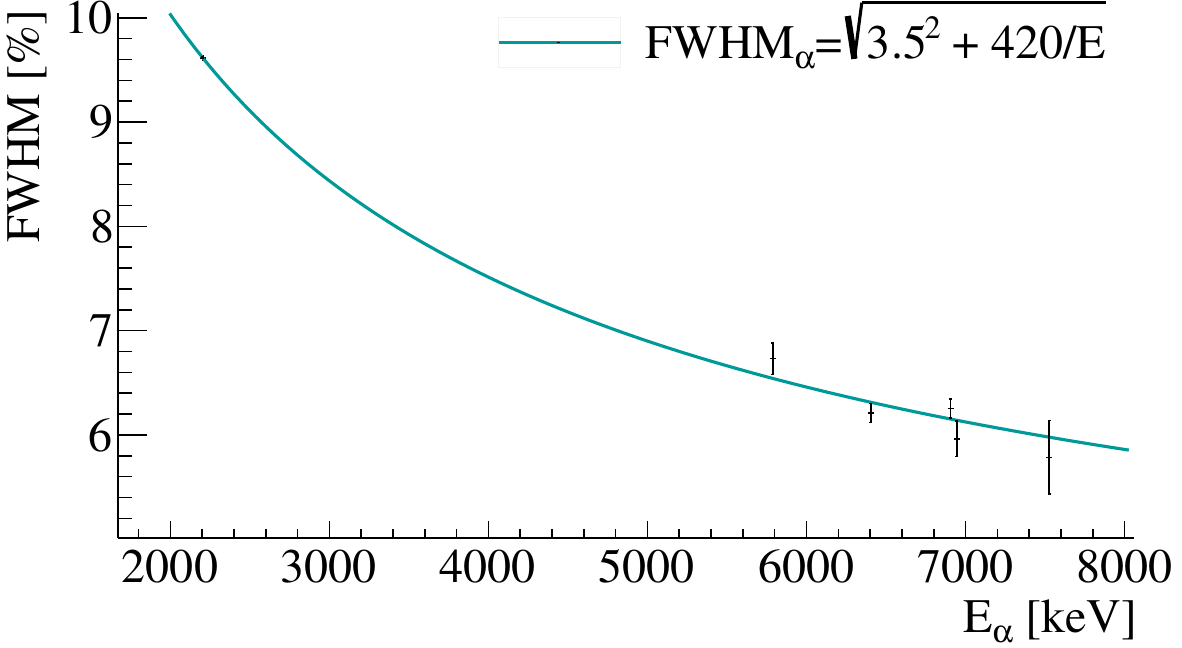}
  \caption{FWHM as a function of energy for \A\ events.}\label{fig:alpha_resolution}
\end{figure}

At this point, we use the identified \A\ peaks to calibrate the \A\ energy scale.
Specifically, we fit the $I$ distribution of each identified \A\ peak with a Gaussian function,
whose position is related  to the $Q$-value of the decay,
while the FWHM of the peak is a measure of the energy resolution.
We include all the \A\ from both triplet decays except that of $^{223}$Ra,
since in this case \G\ rays are also emitted with a probability $>$90\%, introducing additional features to the $I$ distribution.
Figure~\ref{fig:alpha_calib} shows the resulting polynomial fit to the available \A\ calibration points,
including also the one from the $^{152}$Gd decay.
The fit gives us the \A\ energy scale E$_{\alpha}$,
and allows us to extract the resolution curve as a function of energy for \A\ events  (Fig. \ref{fig:alpha_resolution}).

The integral of the exponential fits on the \dt\ distributions corresponds to the number of identified decays.
Taking into account the selection efficiencies of the \A\ peaks,
the duration of the data collection, and the mass of the crystal,
we obtain a (2.34$\pm$0.02)\,mBq/kg contamination value for the $^{224}$Ra$\rightarrow^{220}$Rn$\rightarrow^{216}$Po$\rightarrow^{212}$Pb triplet,
and (1.51$\pm$0.02)\,mBq/kg for the $^{223}$Ra$\rightarrow^{219}$Rn$\rightarrow^{215}$Po$\rightarrow^{211}$Pb one.
As mentioned before, we find no candidate for the two TDC doublets of the $^{238}$U chain.

Considering that all isotopes of the $^{232}$Th chain following $^{228}$Th must be in equilibrium,
we cross-check the obtained activity by searching for the $^{212}$Bi$\rightarrow^{212}$Po$\rightarrow^{208}$Po \B-\A\ TDC (Tab.~\ref{tab:dc}).
Given the shorter half-life value of this TDC, we acquired a special run with a short acquisition window of 400\,ns.
To identify the events, we start from the 8.9\,MeV \A\ of the second decay
and look backward in time to find the preceding \B\ decay.
When calculating the activity with this \B-\A\ coincidence, two efficiencies have to be taken into account:
the first is introduced by the acquisition window of 400\,ns and the digitizer dead time,
which reduces the visible \dt\ distribution to $\sim35\%$, 
and the second  is the 64\% branching ratio.
The resulting activity is (1.9$\pm$0.08)\,mBq/kg. 
This is 20\% smaller than the activity measured with the \A\ triplet,
probably as a result of a poorly measured dead-time efficiency.
Therefore, we consider the value measured with the \A\ triplet as more reliable.

\begin{figure}[htbp]
  \centering
  \includegraphics[width=\columnwidth]{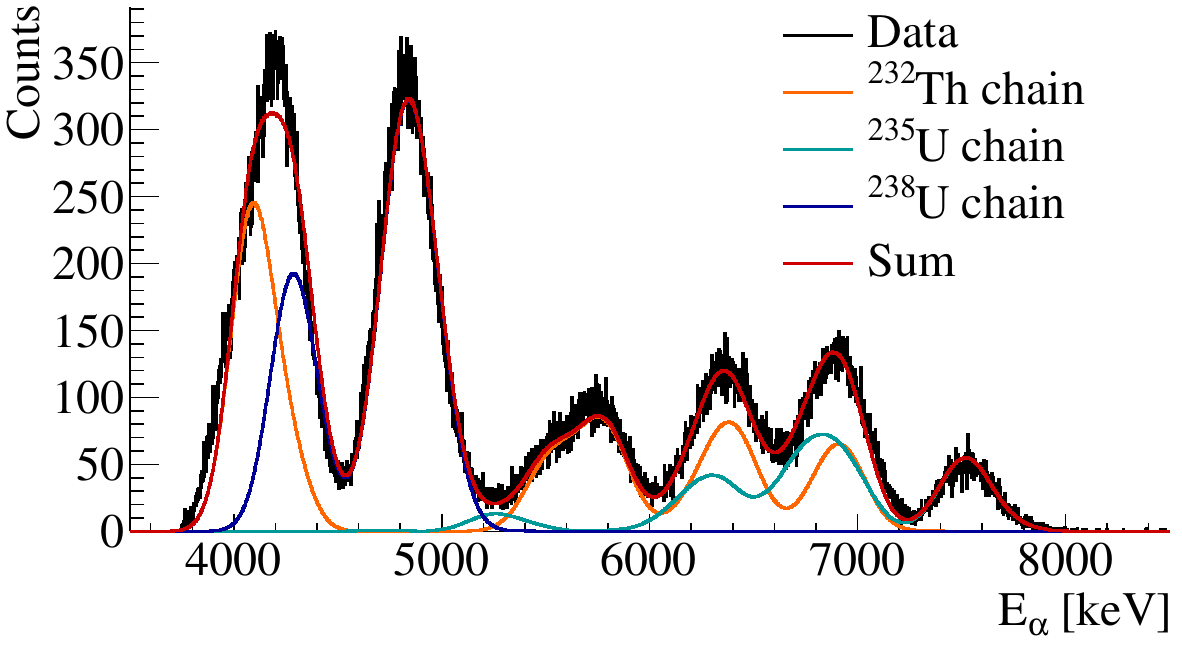}
  \caption{Global fit (red line) to the \A\ peaks (black points).
    The contribution from the individual decay chains are also reported.}\label{fig:alpha_tot_fit}
\end{figure}

Considering a different approach with respect to the TDC study,
we fit the \A\ energy spectrum with an analytical model which includes all possible \A\ peaks
of the three radioactive chains $^{232}$Th, $^{238}$U, and $^{235}$U.
The expected peak from each \A\ decay is modeled with a Gaussian having mean centered on the $Q$-value of the decay
and a FWHM from the fit in Fig. \ref{fig:alpha_resolution}.
In some decays the daughter nucleus can be produced in different excited states
and de-excites by emitting one or more \G\ rays.
In this case we construct a Gaussian for all the possible \A\ energies weighted with their specific probability,
and take into account the energy deposited by \G, constructing further Gaussian distributions
using the calibration curve and energy resolution for the \BG\ events.
This is an approximate procedure, as the case in which the \G\ energy is not fully contained in the crystal is not considered.
We fix the amplitudes of all peaks following $^{228}$Th to 2.34\,mBq/kg,
and of those following $^{227}$Ac to 1.51\,mBq/kg.
On the contrary, we keep the amplitudes of peaks in the early parts of the chains as free parameters.
The fit is performed in the E$_{\alpha}$ space and the result is shown in Fig.~\ref{fig:alpha_tot_fit}.
Tab.~\ref{tab:cont_measure} summarizes the contamination values for the various part of the decay chains
obtained with the TDC and the spectra fit approaches.

An activity of 8.4\,mBq/kg is found for $^{232}$Th \A\ decay, a factor of 3.6 higher than the activity of the rest of the chain. The \A\ decays in the higher part of the $^{238}$U chain show an activity of 6.3\,mBq/kg, while the lower part of the chain is not found in our analysis. For the $^{235}$U chain, \A\ decays of $^{235}$U and $^{231}$Pa are present with an activity of 0.30\,mBq/kg. It is interesting to notice that this activity matches what expected based on the known isotopic ratio of $^{235}$U/$^{238}$U=0.7\% and on the lifetimes of the two isotopes. The lower part of the $^{235}$U chain features a higher activity according to the measurements with the TDC analysis, pinpointing a contamination of $^{227}$Ac. 
The total \A\ rate which is obtained by summing up the contributions from the three radioactive chains matches the total measured \A\ activity above the $^{152}$Gd peak.

This measured background can be compared with the one obtained in similar studies of GAGG crystals~\cite{Omori:2024wvc,Omori:2024zmq}.
As for the \A\ background, it appears that the measured contamination values are comparable for the $^{232}$Th chain.
The $^{238}$U upper part contamination appears a factor of 20 lower in the present study.
As for the $^{235}$U chain, the contamination in the upper part is a factor of 10 lower in our crystal,
while it is comparable in the lower part, as a result of a $^{227}$Ac contamination.

\begin{table*}[htbp]
  \centering
  \caption{Activities of \A\ decays for the various parts of the $^{232}$Th, $^{238}$U and $^{235}$U decay chains.
    The values marked with an asterisk are fixed according to the activity values from the TDC triplets.
    The total calculated \A\ rates for each chain and for the sum of the three chains are also reported.}\label{tab:cont_measure}
  \begin{tabular}{lccc}
    \toprule
    \midrule
    Radioactive chain  &  &  Total \A\ rate  [mHz/kg]\\
    $^{232}$Th &  &  20.1$\pm$0.1 \\
     \midrule
     Decay  &  Q-value [keV]  & Method  &  Activity [mBq/kg] \\
     \midrule
     $^{232}$Th  & 4082 &  \A\ spectrum fit  &  8.4$\pm$0.1 \\
     $^{228}$Th  &  5520 &  \A\ spectrum fit  &  2.34$^{*}$  \\
     $^{224}$Ra$\rightarrow^{220}$Rn$\rightarrow^{216}$Po  & 5789, 6405, 6906 &  triple \A\ coincidence  &   2.34$\pm$0.02  \\
     $^{212}$Bi  & 6207 &  \A\ spectrum fit  &  2.34$\times$0.36\,=\,0.84$^{*}$  \\
     $^{212}$Po & 8954 &  \A\ spectrum fit  &  2.34$\times$0.64\,=\,1.50$^{*}$  \\
     $^{212}$Bi$\rightarrow^{212}$Po  & 2252, 8954 &  \B-\A\ coincidence &  1.22$\pm$0.04  \\
    \midrule
    \midrule
     Radioactive chain  &  &  Total \A\ rate  [mHz/kg]\\
     $^{238}$U &   &  18.9$\pm$0.2 \\
     \midrule
     Decay  &  Q-value [keV]  & Method  &  Activity [mBq/kg] \\
     \midrule
     $^{238}$U  & 4270 &  \A\ spectrum fit  &  6.3$\pm$0.1 \\
     $^{234}$U  & 4857 &  \A\ spectrum fit  &  6.3$\pm$0.1 \\
     $^{230}$Th & 4770 &  \A\ spectrum fit  &  6.3$\pm$0.1 \\
     $^{226}$Ra  & 4871 & \A\ spectrum fit  &  Not found \\
     $^{222}$Rn$\rightarrow^{218}$Po  & 5590, 6115 &  double \A\ coincidence  &  Not found  \\
     $^{214}$Bi$\rightarrow^{214}$Po  & 3269, 7834 & \B-\A\ coincidence  &  Not found \\
     $^{210}$Po  & 5407 &   \A\ spectrum fit  &  Not found \\
     \midrule
     \midrule
     Radioactive chain  &  &  Total \A\ rate  [mHz/kg]\\
     $^{235}$U &   &  8.2$\pm$0.1 \\
     \midrule
     Decay  &  Q-value [keV]  & Method  &  Activity [mBq/kg] \\
     \midrule
     $^{235}$U  & 4678 &  \A\ spectrum fit  &  0.30$\pm$0.05 \\
     $^{231}$Pa  & 5150 &  \A\ spectrum fit  &  0.30$\pm$0.05 \\
     $^{227}$Th  & 6146 &  \A\ spectrum fit  &  1.51$^{*}$ \\
     $^{223}$Ra$\rightarrow^{219}$Rn$\rightarrow^{215}$Po   & 5979, 6946, 7526  &  triple \A\ coincidence  &  1.51$\pm$0.02 \\
     $^{211}$Bi  & 6750 &  \A\ spectrum fit  &  1.51$^{*}$ \\
    \midrule
    \midrule
     Total \A\ rate ($^{232}$Th+$^{238}$U+$^{235}$U) &   &  (47.2$\pm$0.2) mHz/kg \\
     \bottomrule
  \end{tabular}
\end{table*}

\section{Perspectives as neutron detector}\label{sec:neutrons}

\begin{figure}
  \centering
  \includegraphics[width=\columnwidth]{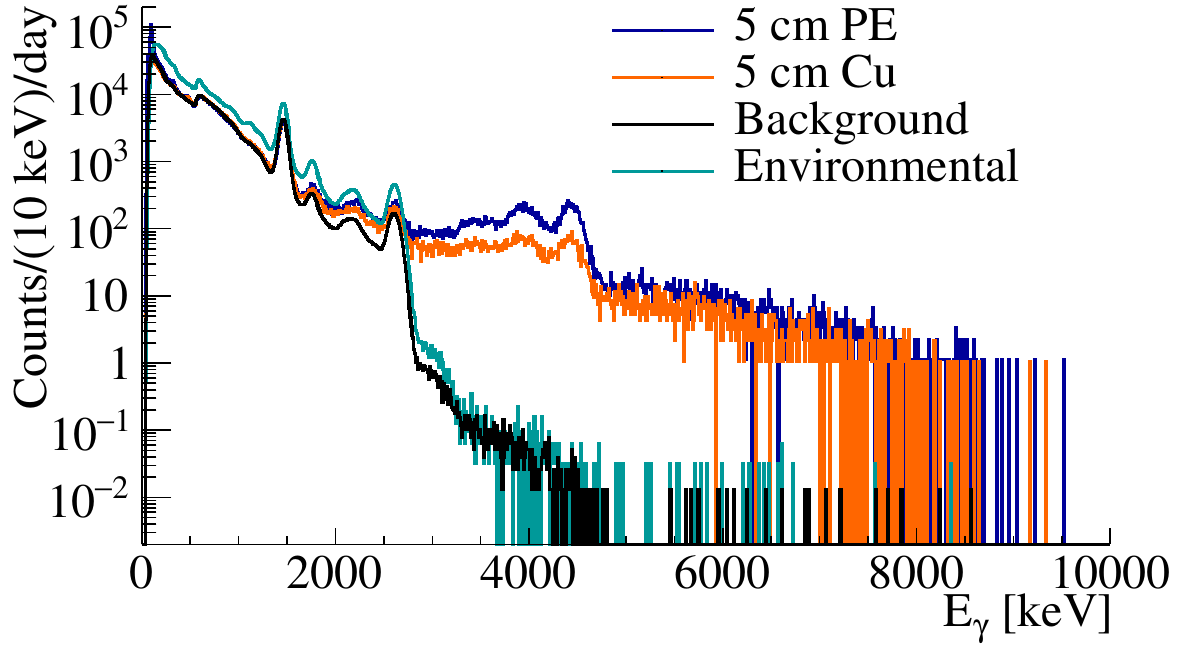}
  \caption{\BG\ spectra obtained with the AmBe calibration and environmental neutron measurements.
    The AmBe spectra in blue and red are obtained
    from the two measurements with a PE moderator and a copper brick placed between the source and the detector, respectively.
    The background spectrum is measured with the detector shielded by the borated PE,
    while the environmental refers to the thermal neutron measurement with no shield.}\label{fig:AmBeThermal}
\end{figure}

In order to demonstrate the possibility of using GAGG crystals for neutron detection underground,
we performed two AmBe calibrations, placing the source inside the borated PE shielding.
In the first measurement, we placed a 5\,cm thick PE moderator between the source and the detector,
then we substituted the PE moderator with a 5\,cm thick copper brick.
Fig.~\ref{fig:AmBeThermal} shows the corresponding spectra in the \BG\ band:
in all cases, a continuum reaching $\sim9$\,MeV is clearly visible,
consisting of a combination of electrons and \G\ emitted from the de-excitation of $^{156}$Gd and $^{158}$Gd
nuclei following a neutron capture on $^{155}$Gd and $^{157}$Gd, respectively.
To our knowledge, this is the first measurement of the high-energy Gd \BG\ cascades with a GAGG crystal.
The spectrum also shows a peak at $4.4$\,MeV, corresponding to a \G\ emitted by the de-excitation of $^{12}$C
produced by the ($\alpha$,n) reaction in the AmBe source, as well as its single and double-escape peaks.

To estimate the relative intensity of the Gd cascades and $^{12}$C-induced events,
we substituted the PE moderator with the copper brick, as previously mentioned,
keeping the distance between the source and the detector unaltered.
This leads to a reduction of a factor 1.6 of the continuum above 5\,MeV,
caused by the absence of the PE moderator, and of a factor 2.4 in the region
containing the 4.4\,MeV peak and its escape peaks, caused by the copper presence.
On the contrary, the two spectra match very well below $\sim$2.7\,MeV,
showing that the neutron contribution is negligible in comparison to the intrinsic background in that energy range.
At present, we do not yet have a thorough evaluation of the overall detector efficiency.
This is only achievable via Monte Carlo simulations, which are currently under development.

Finally, we acquired one month of data with the unshielded detector to measure the thermal neutron flux.
The corresponding spectrum is shown in teal in Fig.~\ref{fig:AmBeThermal},
showing an overall higher count rate than in the background spectrum,
which has different causes, depending on the energy region.
The absence of the borated PE shielding causes a higher rate in the environmental \G\
which dominates the spectrum up to the 2615\,keV line from \Tl,
and also causes an increase in event rate up to $\sim$3.2\,MeV,
due to  the higher probability for the 2615\,keV and the 583\,keV lines,
which are in true coincidence in 85\% of the cases, to be both fully absorbed by the detector.
On the other hand, the higher rate above 3.2\,MeV can only be attributed to neutron-induced events.

Tab.~\ref{tab:thermal} reports the count rates obtained from the measurements with the shielded and unshielded detector
over different energy ranges. The region above 5\,MeV is only populated by neutron-induced events,
but suffers from a limited containment efficiency for the Gd de-excitation \G's.
Expanding the counting region to lower energies allows to include a higher fraction of neutron events,
at the cost of accepting also a higher background.
Converting the net neutron event rates to flux values, we obtain estimates in the \mbox{$[0.5-1]\cdot10^{-6}$\,neutrons/cm$^2$/s},
in agreement with previous measurements of thermal neutrons performed at LNGS~\cite{Best:2015yma,Belli:1989wz,Wulandari:2003cr}.
In the future, we plan to optimize the counting region by means of dedicated simulations.

\begin{table}[htbp]
  \centering
  \caption{Count rates obtained from the backgroud measurement
    and the environmental neutron measurement in different energy ranges.
    The last column reports the net count-rate difference
    between the environmental and background rates, attributed to neutron-induced events.}\label{tab:thermal}
  \begin{tabular}{cccc}
    \toprule
    $\Delta$E & Background & Environmental & Net neutrons \\
    $[$MeV$]$     & $[$counts/day$]$    & $[$counts/day$]$       & $[$counts/day$]$ \\
    \midrule
    $[3.5,10]$ & $5.1\pm0.3$   & $6.3\pm0.4$ & $1.1\pm0.5$ \\
    $[4.0,10]$ & $1.7\pm0.2$   & $2.9\pm0.3$ & $1.2\pm0.3$ \\
    $[4.5,10]$ & $0.45\pm0.08$ & $1.0\pm0.2$ & $0.6\pm0.2$ \\
    $[5.0,10]$ & $0.28\pm0.06$ & $0.7\pm0.2$ & $0.4\pm0.2$ \\
    \bottomrule
  \end{tabular}
\end{table}

\section{Conclusions and outlook}\label{sec:conclusion}

In this article, we have demonstrated the possibility to effectively employ large volume GAGG crystals
for the measurement of neutrons, thanks to the detection of the electron and \G\ cascades 
emitted by the $^{156}$Gd and $^{158}$Gd de-excitation.
An advanced frequency-based PSD algorithm has allowed us to fully separate the \A\ from the \BG\ events.
Applying a delayed-coincidence analysis to the \A\ spectrum,
we have measured with $\leq20\%$ precision the contamination values of
all parts of the \Th, \U, and $^{235}$U decay chains.
Dedicated AmBe calibrations have confirmed the GAGG potential for neutron detection,
and the preliminary measurement of environmental neutrons in the Hall A of LNGS
is in good agreement with previous results obtained with different detectors in the same location.
The main background is represented by the intrinsic \Th\ contamination,
which induces a continuum in the \BG\ band reaching 5\,MeV.
Reducing the \Th\ contamination below the $0.1$\,mBq/kg
is crucial for enhancing the detector sensitivity to neutrons.

In the near future, we aim to develop a full simulation of the neutron interaction in the detector,
and of the propagation of the electron and \G\ rays emitted by the Gd de-excitation following a neutron capture.
Such simulations will allow us to evaluate the total detection efficiency,
and to design a larger setup consisting of several detectors coupled to Bonner spheres
for measuring the neutron spectrum at LNGS and in other underground laboratories.

As for using the GAGG detectors above ground to measure neutrons, the fundamental limitation is the
impossibility to distinguish the neutron-induced \BG\ events from the overwhelming muon background.
In the long term, we intend to bypass this limitation by inserting a GAGG crystal
into a plastic-scintillator case with a thickness of several centimeters, which would act both as a neutron moderator and a muon veto.
Such a design could allow to effectively operate GAGG crystals for above-ground neutron detection,
and provide a valid and affordable alternative to $^{3}$He counters.


\backmatter

\bmhead{Acknowledgements}

This work was supported by the EU Horizon2020 research and innovation
program under the Marie Sklodowska-Curie Grant Agreement No. 754496,
and by the European Union, Next Generation EU, Mission 4 Component 1,
CUP 2022WWRZZP\_001.

\bibliography{bibliography}

\end{document}